# Further Evidence for Intrinsic Redshifts in Normal Spiral Galaxies

David G. Russell
Owego Free Academy
Owego, NY 13827
USA

Abstract

Evidence from galaxy absolute magnitudes, linear diameters, and HyperLeda images is presented which strongly supports the interpretation that some normal spiral galaxies can contain large non-cosmological (intrinsic) redshifts in excess of 5000 km s$^{-1}$.

## 1. Introduction

In a previous analysis (Russell 2005 – hereafter paper 1) evidence was presented that normal spiral galaxies may contain a component of non-velocity (intrinsic) redshift. The results of paper 1 support the previous claims of Arp (1988,1990) that late type (Sbc/Sc) spiral galaxies can have large excess redshift relative to early type spiral galaxies (Sa/Sb) in galaxy clusters. Galaxies of morphology similar to ScI were found to have a systematic excess redshift relative to the predictions of a smooth Hubble flow where the Hubble Constant ($H_0$) is 72 km s$^{-1}$ Mpc$^{-1}$. This excess was identified by comparing the type dependent Tully-Fisher Relation distances (TD-TFR – Russell 2004,2005) with Hubble distances for a sample of ScI galaxies.

In paper 1 it was demonstrated that in addition to the systematic excess redshift, many of the ScI galaxies had differences between TD-TFR and Hubble distances that are too large to be the result of peculiar motions or TFR data errors. In the most extreme cases, distance modulus errors in excess of 1.00 magnitude or peculiar motions in excess of +3500 km s$^{-1}$ would be required to account for the discrepancy between Hubble and TFR distances. Companion galaxies to these extreme cases were identified in paper 1 and found to have TD-TFR distance moduli in agreement with the ScI's – thus supporting the accuracy of the TD-TFR distances of the ScI's.

A number of possible mainstream explanations for the anomalous redshift results were considered in paper 1. Each mainstream explanation was shown to be insufficient to account for the extreme cases of excess redshift. The central issue remains the accuracy of the TD-TFR distances. While evidence from TFR data uncertainty, intrinsic TD-TFR scatter, and companion galaxies provide strong support for the accuracy of the TD-TFR distances additional evidence is needed.

It is important to note that the evidence for intrinsic redshifts in spiral galaxies can be evaluated independent of cosmological models. It has been suggested by Arp (1998) that large intrinsic redshifts require a non-expanding universe. However, Bell (2002,2004) has argued that intrinsic redshifts are superimposed upon expansion of the universe. The evidence discussed in this paper is compatible with either interpretation.

## 2. ScI galaxies with the largest redshift anomalies

In Table I the nine ScI galaxies from paper 1 with the most significant discrepancies between TD-TFR and Hubble distances are listed. Column 1 is the galaxy identification from the ESO catalog; column 2 is the morphological type in HyperLeda (Paturel et al 2003); column 3 is the logarithm of the rotational velocity from Mathewson&Ford (1996 – hereafter MF96); column 4 is the corrected B-band magnitude from HyperLeda; column 5 is the inclination from MF96; column 7 is the redshift corrected to the cosmic microwave background reference frame from HyperLeda; column 8 is the peculiar velocity required to account for the discrepancy between Vcmb observed and Vcmb predicted at the TD-TFR distance if $H_0$=72 km s$^{-1}$ Mpc$^{-1}$.

*Table I: ScI galaxies with large excess redshifts*

| 1 Galaxy (ESO) | 2 Type | 3 Log Vrot | 4 btc | 5 Incl. | 6 Mpc TD-TFR | 7 Vcmb | 8 PV72 |
|---|---|---|---|---|---|---|---|
| 254-22 | ScII | 2.377 | 13.109 | 57 | 77.3 | 10813 | +5247 |
| 445-27 | SBcI-II | 2.389 | 13.363 | 60 | 89.5 | 11722 | +5278 |
| 501-75 | ScII | 2.250 | 12.537 | 61 | 44.1 | 5563 | +2388 |
| 384-9 | SBcII | 2.260 | 13.722 | 59 | 78.0 | 11487 | +5871 |
| 147-5 | SBcI-II | 2.274 | 13.846 | 34 | 85.1 | 10666 | +4539 |
| 297-36 | ScII | 2.274 | 14.139 | 56 | 97.7 | 10972 | +3938 |
| 30-14 | ScI-II | 2.228 | 13.536 | 31 | 66.4 | 8207 | +3426 |
| 52-20 | SBbcI-II | 2.196 | 13.590 | 63 | 63.2 | 8068 | +3518 |
| 545-13 | ScI-II | 2.350 | 13.520 | 34 | 87.5 | 9923 | +3623 |

The galaxies in Table I have excess redshifts ranging from +2388 km s$^{-1}$ to +5871 km s$^{-1}$. Since real peculiar motions are not expected to exceed 1500 km s$^{-1}$ in galaxy clusters (e.g. Karachentsev et al 2003), these excess redshifts were interpreted in paper 1 as evidence for intrinsic redshifts. It is important to note that these galaxies represent the extreme examples of a systematic trend of excess redshift relative to the predictions of a smooth Hubble flow (see Fig. 2 of paper 1). Their importance is that the redshift deviations are too large to be accounted for by real peculiar motions, intrinsic TFR scatter, or TFR data errors.

Throughout this paper excess redshift values will be quoted relative to the currently preferred Hubble Constant of 72 km s$^{-1}$ Mpc$^{-1}$ (Freedman et al 2001) unless stated otherwise. In paper 1 it was suggested that a Hubble Constant of 55-60 km s$^{-1}$ Mpc$^{-1}$ might be preferred after accounting for intrinsic redshifts. While it would be appropriate to utilize this lower value of the Hubble Constant for estimates of the size of intrinsic redshifts, $H_0$=72 km s$^{-1}$ Mpc$^{-1}$ is utilized in this paper for the purpose of constraining the intrinsic redshifts to their conservative lower limits. Should a lower value of the Hubble Constant - such as 60 km s$^{-1}$ Mpc$^{-1}$ - be adopted, the intrinsic redshift values quoted here and in paper 1 would need to be increased.

In the sections that follow further evidence is presented that supports the accuracy of the TD-TFR distances and the interpretation that some normal galaxies possess large intrinsic redshifts. Section 3 discusses the TFR errors required to accommodate these redshift discrepancies. Section 4 discusses evidence from linear diameters. Section 5 presents additional evidence from HyperLeda images.

## 3. Evidence that the excess redshifts are not caused by large TFR errors

Table II provides the TFR data errors required to account for the difference between the redshift and TD-TFR distances of the nine ScI group galaxies in Table I. Column 1 is the galaxy identification; column 2 is the TD-TFR distance modulus; column 3 is the uncertainty in the TD-TFR distance modulus derived from errors in B-band magnitude and rotational velocity provided in HyperLeda; column 4 is the Hubble distance modulus ($H_0$=72 km s$^{-1}$ Mpc$^{-1}$); column 5 is the difference between the Hubble and the TD-TFR distance moduli; column 6 is the rotational velocity from MF96; column 7 is the rotational velocity required for the TD-TFR distance modulus to equal the Hubble distance modulus; column 8 is the absolute magnitude of the galaxy at the TD-TFR distance; column 9 is the absolute magnitude of the galaxy at the Hubble distance; column 10 is the inclination required for the galaxy to have the rotational velocity in column 7 assuming a rotational velocity error caused by inclination error.

The inclinations in column 10 are closer to face-on orientation than the observed inclinations and therefore the internal extinction correction to the B-band magnitude will be smaller than actually applied in HyperLeda. The rotational velocities in column 7 are determined after adjusting the extinction correction to the B-band magnitude to account for the alternate inclination. This adjustment to the extinction correction amounts to ~0.30 mag for ESO 254-22, ESO 445-27, ESO 501-75, ESO 384-9, ESO 297-36, and ESO 52-20. Since ESO 147-5, ESO 30-14, and ESO 545-13 have low inclinations (<35 degrees) and therefore small internal extinction corrections, no adjustment was made to the HyperLeda magnitudes for these 3 galaxies. It should be noted that if the rotational velocity is assumed to be incorrect for reasons other than inclination errors, then the rotational velocities in column 7 would need to be increased.

*Table II: Required TF errors*

| 1 | 2 | 3 | 4 | 5 | 6 | 7 | 8 | 9 | 10 |
|---|---|---|---|---|---|---|---|---|---|
| Galaxy | m-M TD-TFR | +/- | m-M$_{72}$ | $\Delta$m-M | Vrot | Vrot$_{72}$ | M$_{TF}$ | M$_{72}$ | Incl$_{72}$ |
| 254-22 | 34.44 | .35 | 35.88 | 1.44 | 238 | 400 | -21.53 | -22.77 | 30 |
| 445-27 | 34.76 | .29 | 36.06 | 1.30 | 245 | 370 | -21.40 | -22.69 | 32 |
| 501-75 | 33.22 | .17 | 34.44 | 1.22 | 178 | 259 | -20.68 | -21.90 | 37 |
| 384-9 | 34.46 | .32 | 36.01 | 1.55 | 182 | 312 | -20.74 | -22.29 | 30 |
| 147-5 | 34.65 | .91 | 35.85 | 1.20 | 188 | 323 | -20.80 | -22.00 | 19 |
| 297-36 | 34.95 | .23 | 35.91 | .96 | 188 | 259 | -20.81 | -21.77 | 37 |
| 30-14 | 34.11 | .54 | 35.28 | 1.17 | 169 | 286 | -20.57 | -21.74 | 18 |
| 52-20 | 34.00 | .18 | 35.25 | 1.25 | 157 | 233 | -20.41 | -21.66 | 37 |
| 545-13 | 34.71 | .47 | 35.70 | .99 | 224 | 349 | -21.19 | -22.18 | 21 |

### 3.1 TD-TFR distance modulus errors

Column 5 of Table II indicates the TD-TFR distance modulus error if the Hubble distances are correct. The errors would range from 0.96 mag to 1.55 mag and are significantly larger than the uncertainty from data errors provided in column 3. In addition, the observed scatter of the TD-TFR was found to be only +/-0.22 mag (Russell 2004) and therefore the required errors are also too large to be accounted for by intrinsic TFR scatter.

### 3.2 Absolute magnitudes

Columns 8 and 9 of Table II give the absolute magnitudes at the TD-TFR distance and the Hubble distance respectively. Among the ScI group calibrators (paper 1) absolute magnitudes range from –20.44 to –21.53 (Table III). The absolute magnitudes of the nine galaxies in Table II would range from –21.66 to –22.77 at the Hubble distances. Therefore each of these galaxies would be more luminous than the most luminous ScI group calibrator (NGC 7331 – Table III). Figure 1 illustrates how extreme these absolute magnitudes would be at a given rotational velocity relative to the mean relation defined by the calibrator sample. For example ESO 254-22 and ESO 445-27 have M$_{72}$ of –22.77 and –22.69 respectively. This is comparable with the lowest luminosity quasars (M=-23).

The absolute magnitudes at the Hubble distances are also very large compared with the most luminous elliptical and CD galaxies in the SBF survey of Tonry et al (2001-Table IV). These comparisons illustrate that if the Hubble distances are correct, the ScI galaxies in Table I have unprecedented luminosities for late type spiral galaxies.

*Table III: ScI calibrators*

| Galaxy | Type | m-M | Reference | Vrot | Log Vrot | $M_B$ |
|--------|------|-----|-----------|------|----------|-------|
| Direct | | | | | | |
| N253 | SBcII | 27.98 | 1 | 196 | 2.292 | -20.99 |
| N1365 | SBbI(SY) | 31.27 | 2 | 249 | 2.397 | -21.37 |
| N1425 | SbII | 31.70 | 2 | 186 | 2.270 | -20.87 |
| N2903 | SBbcI-II | 29.75 | 3 | 192 | 2.282 | -20.89 |
| N3198 | SBcII | 30.70 | 2 | 153 | 2.184 | -20.48 |
| N3627 | SBbII(SY) | 30.01 | 2 | 212 | 2.326 | -21.03 |
| N4258 | SBbcII-III(SY) | 29.51 | 2 | 212 | 2.327 | -21.12 |
| N4321 | SBbcI | 30.91 | 2 | 218 | 2.338 | -21.12 |
| N4535 | SBcI-II | 30.99 | 2 | 186 | 2.270 | -20.64 |
| N4536 | SBbcI-II | 30.87 | 2 | 170 | 2.230 | -20.44 |
| N4603 | SBcI-II | 32.61 | 4 | 228 | 2.358 | -21.26 |
| N7331 | SbcI-II | 30.84 | 2 | 265 | 2.424 | -21.53 |
| | | | | | | |
| Cluster | | | | | | |
| N753 | SBbcI-II | 33.71 | 5 | 208 | 2.317 | -21.18 |
| N818 | ScI | 33.71 | 5 | 233 | 2.366 | -21.31 |
| N536 | SBbI | 33.76 | 5 | 254 | 2.405 | -21.42 |
| N7610 | SBcI-II | 33.65 | 6 | 149 | 2.172 | -20.53 |
| N3095 | SBcII | 32.58 | 6 | 211 | 2.324 | -20.99 |
| IC2560 | SBb(SY) | 32.58 | 6 | 200 | 2.301 | -21.02 |
| N3223 | SBbI-II | 32.58 | 6 | 269 | 2.429 | -21.56 |
| N3347 | SBbI-II | 32.58 | 6 | 213 | 2.327 | -21.17 |
| N3464 | ScII | 33.46 | 2 | 195 | 2.290 | -21.02 |
| 445-58 | SBbcI-II | 33.56 | 2 | 187 | 2.271 | -20.65 |

References: 1 – Karachentsev et al. (2003), 2 – Freedman et al (2001), 3 Drozdovsky and Karachentsev (2000), 4 – Newman et al. (1999), 5 – Prugniel and Simien (1996), 6 – Tonry et al (2001).

*Table IV: CD Galaxies*

| Galaxy | Cluster | m-M SBF | $M_B$ |
|--------|---------|---------|-------|
| N1316 | Fornax | 31.60 | -21.94 |
| N3557 | N3557 | 33.24 | -22.33 |
| N4406 | Virgo | 31.11 | -21.38 |
| N4472 | Virgo | 31.00 | -21.83 |
| N4486 | Virgo | 30.97 | -21.54 |
| N4696 | Centaurus | 32.69 | -21.37 |
| N7619 | Pegasus | 33.56 | -21.71 |

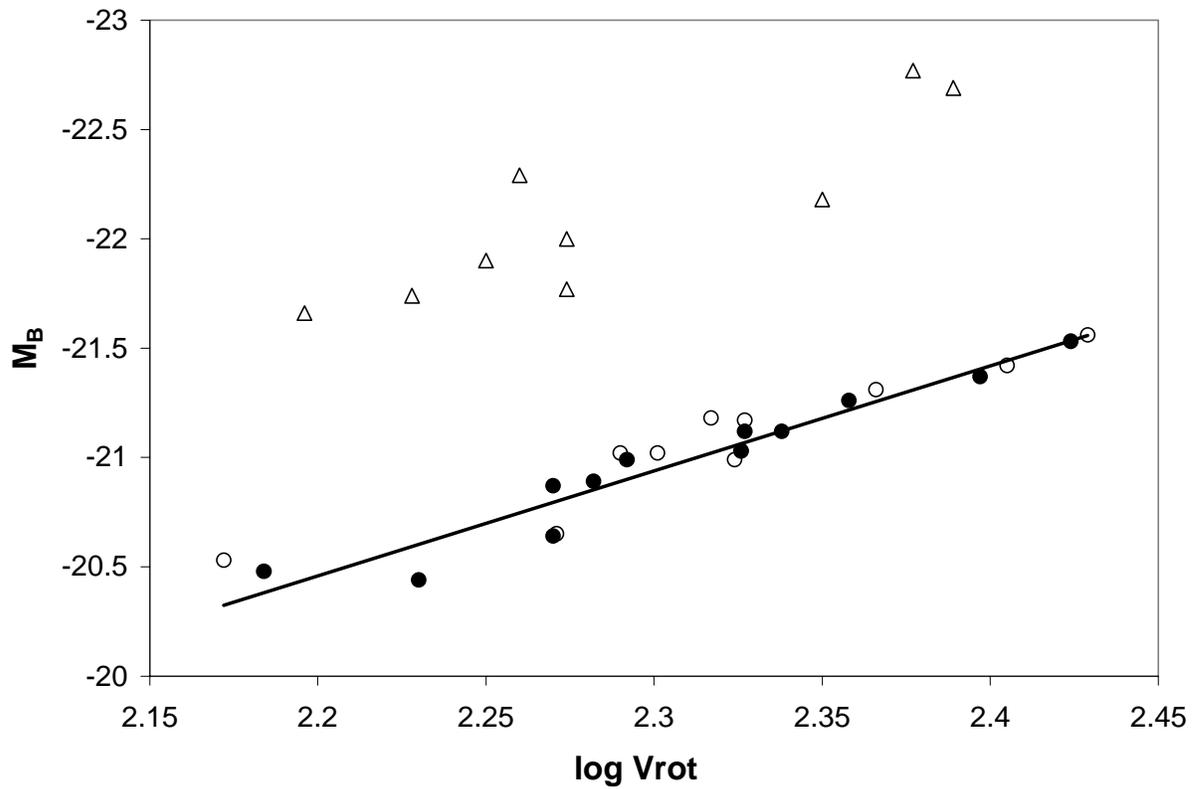

Figure 1: Tully-Fisher plot for calibrators (Table III) and ScI galaxies (Table I). Filled circles are ScI calibrators from paper 1. Open circles are ScI cluster galaxies with cluster distance determined from the surface brightness fluctuation method or Fundamental Plane. Open triangles are the absolute magnitudes of the galaxies in Table I at the Hubble distances.

### 3.3 Rotational Velocities and Inclinations

Column 7 of Table II gives the rotational velocity needed for each galaxy to have a TD-TFR distance modulus equal to the Hubble distance modulus (column 4 of Table II) assuming a rotational velocity error caused entirely by an inclination error. These errors would be from 71 km s$^{-1}$ to 162 km s$^{-1}$ and may be compared with the actual rotational velocities from MF96 given in column 6 and the rotational velocities of the calibrators in Table III. For example ESO 254-22 has an observed rotational velocity of 238 km s$^{-1}$, but requires a rotational velocity of 400 km s$^{-1}$ for the TD-TFR distance to equal the Hubble distance. Note that a rotational velocity of 400 km s$^{-1}$ is 131 km s$^{-1}$ faster than the fastest rotator among the calibrators in Table III.

Column 10 of Table II gives the inclinations required to correct the observed rotational velocities given in column 6 to the rotational velocity given in column 7. ESO 445-27 and ESO 254-22 would need to have inclinations of 27 and 26 degrees respectively compared with 60 degree and 57 degree inclinations given in MF96 – requiring *inclination errors* greater than 30 degrees! Visual inspection of the HyperLeda images of these two galaxies confirms that they are not close to face-on orientation (see figs 8a and 8b).

### 4. Evidence from Linear Diameters

The linear diameters of the galaxies in Table I provide additional evidence for the presence of intrinsic redshifts in these galaxies. Figure 2 is a plot of linear diameter vs. rotational velocity for the ScI sample from paper 1. The sample is restricted to the 89 galaxies with a mean surface brightness in the $D_{25}$ isophote (bri25 in HyperLeda) from 22.80 mag arc sec$^{-2}$ to 23.59 mag arc sec$^{-2}$. Thirteen ScI galaxies from the paper 1 sample fall outside this bri25 range are and therefore not plotted in Fig. 2.

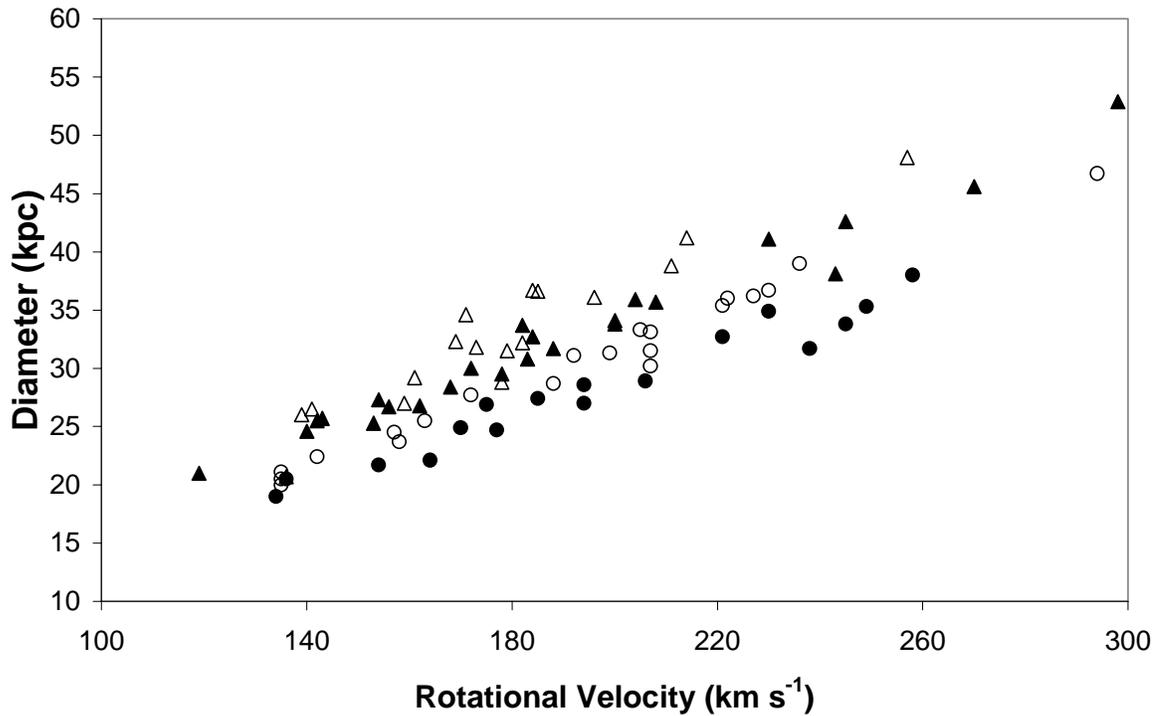

Figure 2 – Linear diameter vs. rotational velocity relationship for 89 ScI galaxies from paper 1 with bri25 from 22.80 mag arc sec$^{-2}$ to 23.59 mag arc sec$^{-2}$. Filled circles are bri25 22.80 to 22.99, open circles are bri25 23.00 to 23.19, filled triangles are bri25 23.20 to 23.39, open triangles are bri25 23.40 to 23.59.

The 89 galaxies were binned into 4 mean surface brightness ranges: bri25 22.80 to 22.99 (filled circles); bri25 23.00 to 23.19(open circles); bri25 23.20-23.39(filled triangles); bri25 23.40-23.59 (open triangles). Fig. 2 demonstrates that at a given rotational velocity, high surface brightness disks have smaller linear diameters than low surface brightness disks. It is important to account for this effect as the difference in linear diameter can vary by over 10 kpc at a given rotational velocity as bri25 varies. Least squares fits were made to each of the 4 surface brightness bins resulting in the following equations:

$$Kpc = 0.1403(Vrot) + 0.64 ; \quad bri25\ 22.80\text{-}22.99 ; \sigma = 1.3\ kpc \qquad (1)$$

$$Kpc = 0.1674(Vrot) - 1.94 ; \quad bri25\ 23.00\text{-}23.19 ; \sigma = 0.9\ kpc \qquad (2)$$

$$Kpc = 0.1724(Vrot) - 0.11 ; \quad bri25\ 23.20\text{ --}23.39; \sigma = 1.5\ kpc \qquad (3)$$

$$Kpc = 0.1940(Vrot) - 1.56 ; \quad bri25\ 23.40\text{-}23.59 ; \sigma = 1.8\ kpc \qquad (4)$$

The expected linear diameter for a galaxy may then be predicted from its rotational velocity and the appropriate equation based upon its bri25 value. Table V provides the diameter information for the nine ScI galaxies and the seven TD-TFR calibrators (Russell 2004,2005) with ScI morphology and bri25 from 22.80 to 23.59. Column 1 is the galaxy identification; column 2 is the rotational velocity, column 3 is bri25 from HyperLeda; column 4 is the $D_{25}$ angular diameter from HyperLeda; column 5 is the linear diameter at the TD-TFR distance or calibrator distance; column 6 is the linear diameter the galaxy would have at its Hubble distance from Table II; column 7 is the linear diameter predicted from equations 1-4; column 8 is the number of standard deviations the Hubble diameter varies from the predicted diameter.

*Table V: Linear Diameters*

| 1 | 2 | 3 | 4 | 5 | 6 | 7 | 8 |
|---|---|---|---|---|---|---|---|
| Galaxy | Vrot | bri25 | Arc min | $Kpc_{TF}$ | $Kpc_{72}$ | $Kpc_{pred}$ | $\sigma_{72}$ |
| 254-22 | 238 | 22.83 | 1.41 | 31.7 | 61.6 | 34.0 | 21.2 |
| 445-27 | 245 | 22.87 | 1.30 | 33.8 | 61.6 | 35.0 | 20.5 |
| 501-75 | 178 | 23.26 | 2.30 | 29.5 | 51.7 | 30.6 | 23.4 |
| 384-9 | 182 | 23.45 | 1.42 | 32.2 | 65.9 | 33.7 | 17.9 |
| 147-5 | 188 | 23.23 | 1.28 | 31.7 | 55.1 | 32.3 | 15.2 |
| 297-36 | 188 | 23.01 | 1.01 | 28.7 | 44.8 | 29.5 | 17.0 |
| 30-14 | 169 | 23.57 | 1.67 | 32.3 | 55.4 | 31.2 | 13.4 |
| 52-20 | 157 | 23.09 | 1.33 | 24.5 | 43.4 | 24.3 | 21.2 |
| 545-13 | 224 | 22.69 | 1.15 | 29.3 | 46.1 | 29.1 | 13.1 |
| | | | | | | | |
| N253 | 196 | 22.89 | 26.30 | 30.2 | | 28.1 | |
| N2903 | 192 | 23.15 | 12.59 | 32.2 | | 30.1 | |
| N3198 | 153 | 23.39 | 7.59 | 30.5 | | 26.3 | |
| N4321 | 218 | 23.02 | 7.59 | 33.9 | | 34.6 | |
| N4535 | 186 | 23.35 | 6.92 | 31.9 | | 32.0 | |
| N4536 | 170 | 23.49 | 7.08 | 30.9 | | 31.4 | |
| N7331 | 265 | 23.53 | 11.22 | 48.3 | | 49.9 | |

Figures 3-6 plot the linear diameters vs. rotational velocity for the four bri25 bins represented by equations 1-4 respectively. Open circles are the linear diameters of the ScI galaxies from paper 1 at the TD-TFR distance and the solid line is the least squares fit. Filled circles are the linear diameters of the TD-TFR calibrators at the calibrator distances. Filled triangles are the linear diameters of the ScI galaxies in Table I at the Hubble distances. It is evident in figures 3-6 that at the calibrator distances, the calibrators fit tightly to the linear diameter relation defined by the ScI sample at the TD-TFR distances.

It can be seen in Table V and figures 3-6 that the linear diameters of the nine ScI galaxies are significantly discrepant at the Hubble distances. For example, ESO 445-27 would have a diameter at the Hubble distance of 61.6 kpc, or $20.5\sigma$ larger than predicted by equation 1. All nine of the galaxies from Table I have linear diameters at least $13\sigma$ too large at the Hubble distances. It is extremely unlikely that the discrepancy between Hubble and TD-TFR diameter results could result from a coupling between diameter and magnitude errors. Errors in corrected magnitudes and angular diameters in HyperLeda are connected through an opacity correction, but this correction is weak for diameters (Theureau 2004, private communication). Therefore, the linear diameter vs. rotational velocity relationship can be treated independent of the magnitude vs. rotational velocity relationship. Since rotational velocity is also measured independent of magnitudes, large errors in magnitude (and therefore the TD-TFR distance) will also result in large errors in the linear diameters. Such large errors are not seen in the tight diameter relations (Figures 3-6) derived from the TD-TFR distances. Therefore the linear diameters provide strong additional support for the accuracy of the TD-TFR distances and the interpretation that these galaxies contain large intrinsic redshifts ranging from $+2388$ km s$^{-1}$ to $+5871$ km s$^{-1}$.

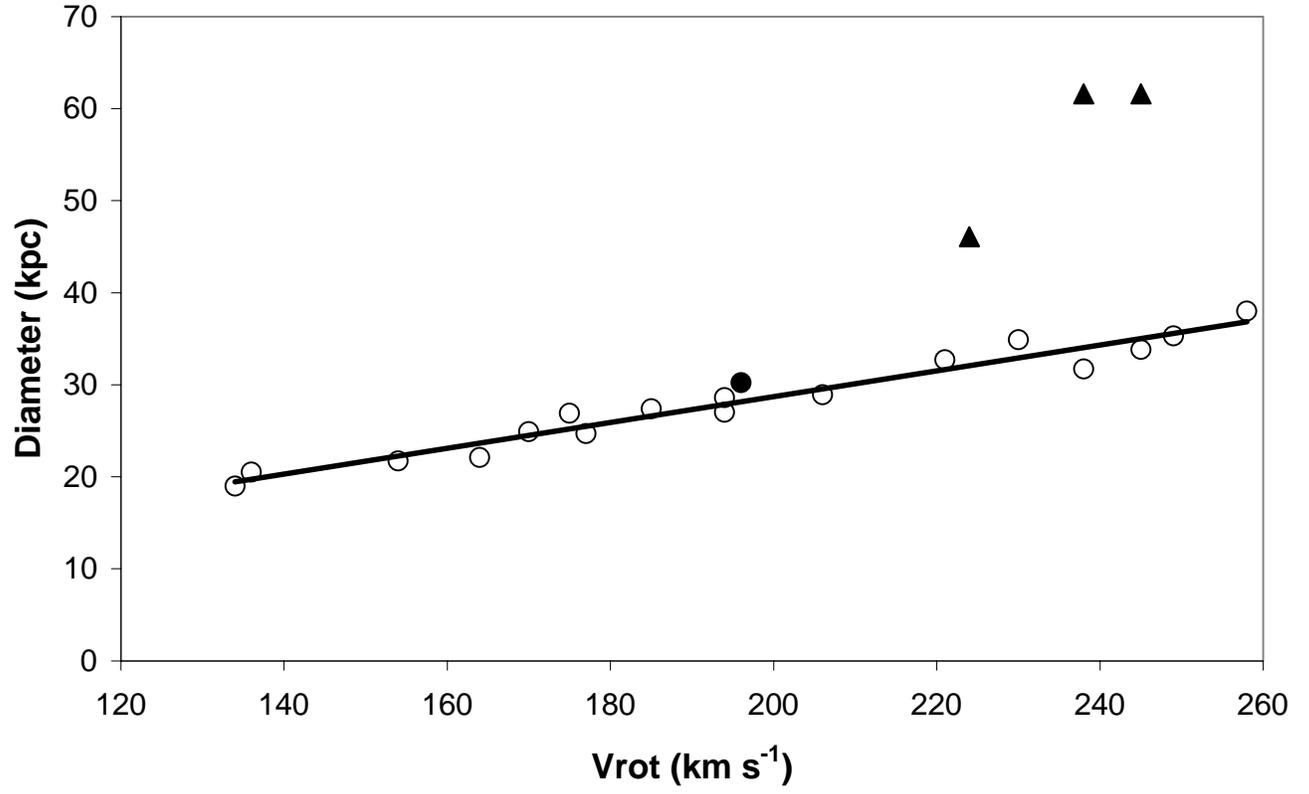

Figure 3 – Linear diameter vs. rotational velocity plot for ScI galaxies with mean surface brightness from 22.80 to 22.99 mag arc sec$^{-2}$.  Filled circle is NGC 253.  Filled triangles are ESO 254-22, ESO 445-27, and ESO 545-13 at the Hubble distances.

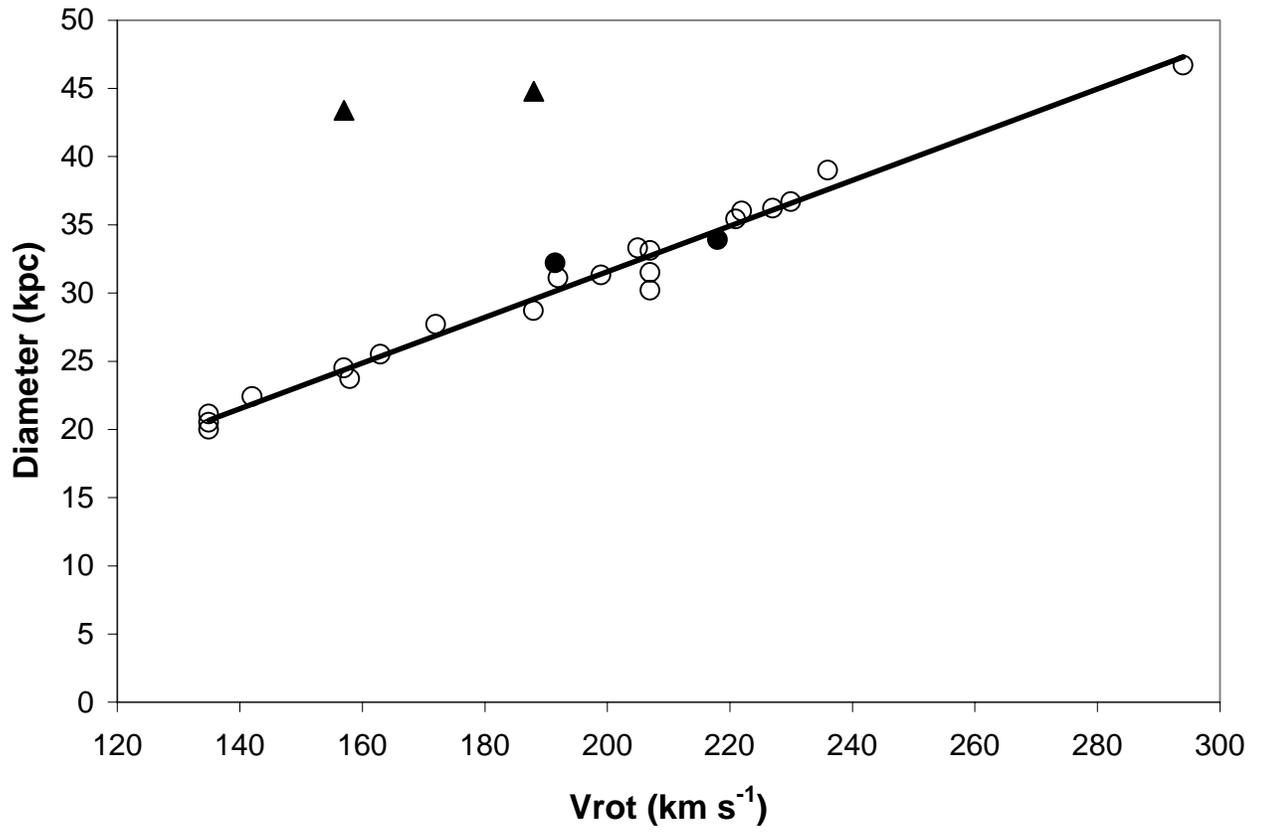

Figure 4 - Linear diameter vs. rotational velocity plot for ScI galaxies with mean surface brightness from 23.00 to 23.19 mag arc sec$^{-2}$. Filled circles are NGC 2903 and NGC 4321  Filled triangles are ESO 52-20 and ESO 297-36 at the Hubble distances.

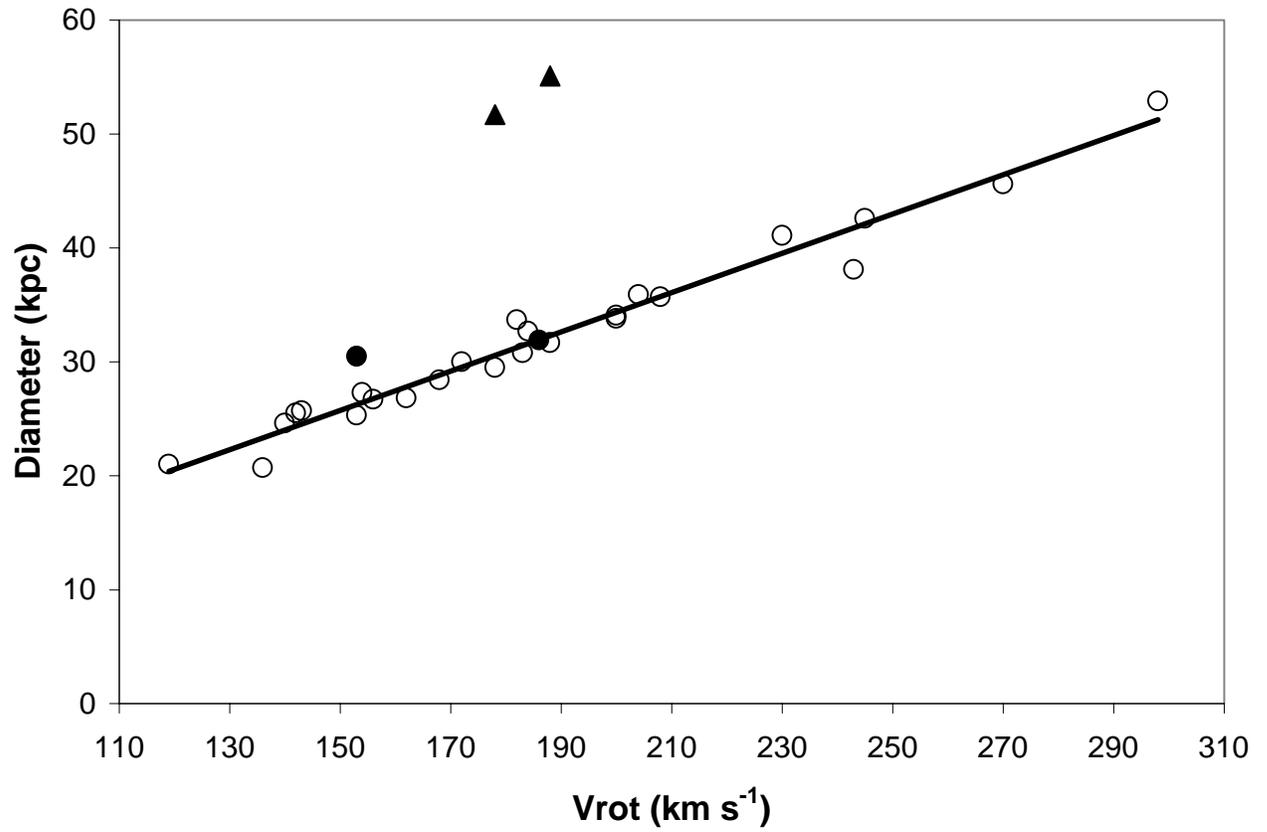

Figure 5 - Linear diameter vs. rotational velocity plot for ScI galaxies with mean surface brightness from 23.20 to 23.39 mag arc sec$^{-2}$.  Filled circles are NGC 3198 and NGC 4535.  Filled triangles are ESO 501-75 and ESO 147-5 at the Hubble distances.

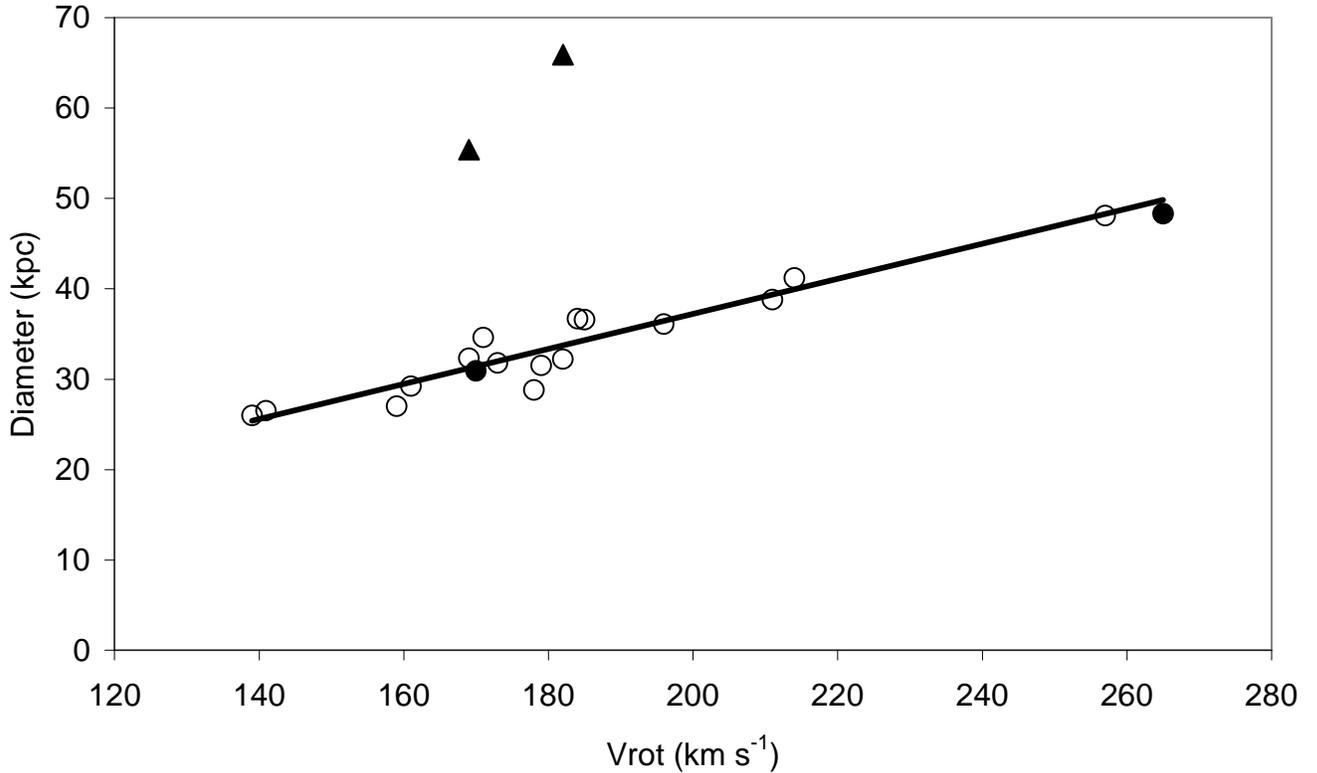

Figure 6 - Linear diameter vs. rotational velocity plot for ScI galaxies with mean surface brightness from 23.40 to 23.59 mag arc sec⁻². Filled circles are NGC 4536 and NGC 7331. Filled triangles are ESO 30-14 and ESO 384-9 at the Hubble distances.

## 5. HyperLeda Images

ESO 5' images from HyperLeda were inspected as an additional check on the interpretation that the galaxies in Table I have large intrinsic redshifts. Examination of the HyperLeda images was guided by the principle that ScI galaxies with nearly the same rotational velocity, mean surface brightness (bri25), and distance should have very similar appearance. This principle is justified by the small scatter in the TD-TFR and the tight relationship between rotational velocity and linear diameters discussed in section 4.

Since the Hubble distances are approximately twice the TD-TFR distances for the galaxies in Table I, visual differences in the HyperLeda images are predicted to be large enough to determine whether the TD-TFR distances or the Hubble distances are more accurate. Comparison galaxies with nearly the same rotational velocity, mean surface brightness, and TD-TFR as the galaxies in Table I were selected from the ScI sample of Russell (2005 – paper 1). However, the comparison galaxies were also required to have close agreement between the Hubble distance ($H_0$=72) and the TD-TFR distance. Thus the comparison galaxies will be galaxies for which the Hubble relation agrees with the Tully-Fisher relation and therefore the distances should not be disputed. The selected comparison galaxies and properties are listed in Table VI with the similar galaxies from Table I.

ESO 186-47 is compared with ESO 52-20 (Table I) in Figures 7a and 7b respectively. Note from Table VI the very close agreement in rotational velocity, surface brightness, and TD-TFR distances for these two galaxies. ESO 52-20 has a redshift of 8069 km s$^{-1}$ compared with 4482 km s$^{-1}$ for ESO 186-47. Thus according to the Hubble relation ESO 52-20 should be at nearly twice the distances of ESO 186-47. Visual inspection of Figures 7a and 7b reveals that the two galaxies have virtually identical appearance as predicted from the TD-TFR distances. This result supports the previous interpretation that ESO 52-20 has an intrinsic redshift of at least +3500 km s$^{-1}$.

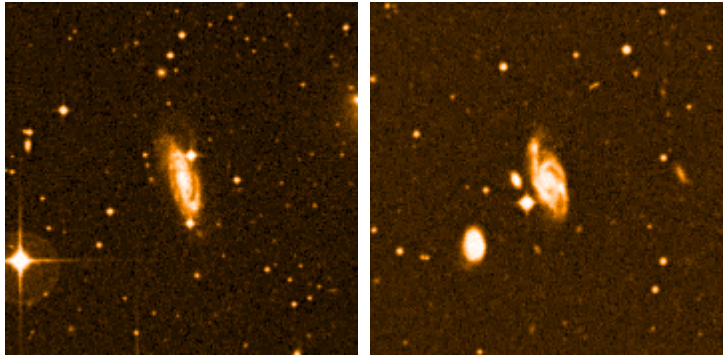

Figure 7a - ESO 186-47        Figure 7b - ESO 52-20

Table VI: Galaxies compared in Figures 7 and 8

| 1 | 2 | 3 | 4 | 5 | 6 | 7 | 8 | 9 |
|---|---|---|---|---|---|---|---|---|
| Galaxy | Vrot | Incl. | bri25 | Mpc$_{TF}$ | kpc | Vcmb | H$_0$ | Mpc$_{72}$ |
| 186-47 | 162 | 62 | 23.21 | 60.3 | 26.8 | 4482 | 74.3 | 62.3 |
| 52-20 | 157 | 63 | 23.09 | 63.2 | 24.5 | 8069 | 127.7 | 112.1 |
| | | | | | | | | |
| 254-22 | 238 | 58 | 22.83 | 77.3 | 31.7 | 10813 | 139.9 | 150.2 |
| 445-27 | 245 | 67 | 22.87 | 89.5 | 33.8 | 11722 | 131.0 | 162.8 |
| 545-13 | 224 | 36 | 22.69 | 87.5 | 29.3 | 9923 | 113.4 | 137.8 |
| 244-21 | 230 | 47 | 22.97 | 103.3 | 34.9 | 6985 | 67.6 | 97.0 |
| P55979 | 235 | 32 | 23.05 | 183.7 | 27.3 | 12378 | 67.4 | 171.9 |

Figures 8a,8b, and 8c are the HyperLeda 5' images of ESO 254-22, ESO 445-27, and ESO 545-13 respectively. All three galaxies have been interpreted to have large intrinsic redshifts and have very similar rotational velocity, mean surface brightness, and TD-TFR distances (Table VI). Visual

inspection of the HyperLeda images of these three galaxies confirms that - as predicted - they have similar appearance.

Two galaxies were selected for comparison with Figures 8a-8c. ESO 244-21 (Figure 8d) and PGC 55979 (Figure 8e) both have rotational velocities and surface brightness close to those of ESO 254-22, ESO 445-27, and ESO 545-13. Both galaxies also have Hubble distances ($H_0$=72 km s$^{-1}$ Mpc$^{-1}$) in close agreement with their TD-TFR distances.

ESO 244-21 has a slighter larger TD-TFR distance than the three galaxies in Figures 8a – 8c. Visual inspection of Fig. 8d confirms that ESO 244-21 has a similar appearance to all three galaxies and thus adds supports the accuracy of the TD-TFR distances.

PGC 55979 from the Serpens cluster sample of paper 1 has a redshift velocity of 12378 km s$^{-1}$ which is only 656 km s$^{-1}$ greater than the redshift velocity of ESO 445-27. Since the redshift distances of PGC 55979 and ESO 445-27 agree within 5.5%, the two galaxies should have a very similar appearance if the redshift distances are correct. Figures 8b and 8e clearly show that PGC 55979 appears to be at a much greater distance than ESO 445-27. This result is exactly as found from the TD-TFR.

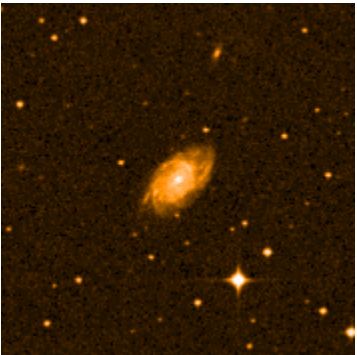 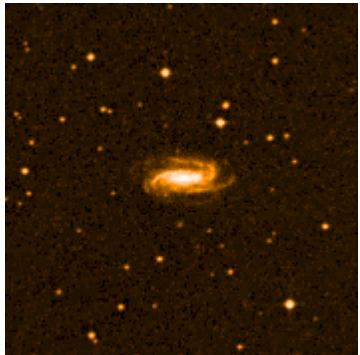 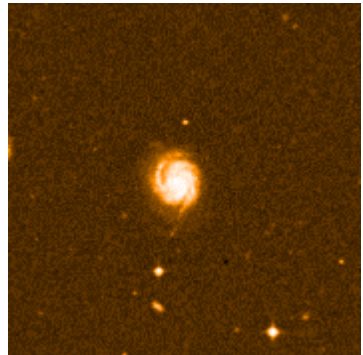

Figure 8a – ESO 254-22     Figure 8b – ESO 445-27     Figure 8c - 545-13

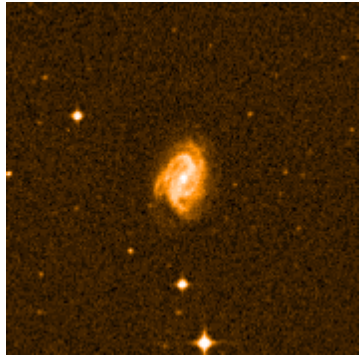 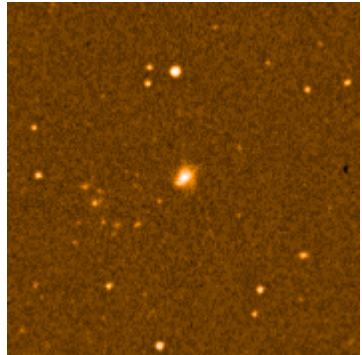

Figure 8d – ESO 244-21     Figure 8e – PGC 55979

Visual inspection of ESO 5' images from HyperLeda confirms the quantitative results presented in this paper and paper 1. If Hubble distances are accurate then large visual differences should be evident between ESO 186-47 and ESO 52-20 and between ESO 445-27, ESO 254-22, ESO 545-13, and ESO 244-21.  The large visual differences are not observed.  Instead the HyperLeda images reveal that when the TD-TFR distances are used, the expected visual similarities and differences are quite evident.  Thus the HyperLeda 5' images provide striking visual  support for the interpretation that the galaxies in Table I contain large intrinsic redshifts.  This result is not limited to the galaxies in Figures 7 and 8.  For example, the interested reader is encouraged to compare the HyperLeda image of ESO 147-5 (Vcmb=10,666 km s$^{-1}$) with the HyperLeda image of the Serpens filament galaxy PGC 55872 (Vcmb=9127 km s$^{-1}$).

## 6.   Conclusion

The results of this analysis provide strong additional support for the presence of large redshift anomalies in the nine galaxies in Table I.  These redshift anomalies have previously been interpreted as evidence of non-cosmological (intrinsic) redshifts (Russell 2005).  The galaxies in Table I are not isolated cases but represent the extreme examples of a trend of excess redshift relative to a smooth Hubble flow found for a sample of 102 ScI galaxies (Russell 2005).   TD-TFR distances, limits on TD-TFR scatter, companion galaxy distances, absolute magnitudes, linear diameters, and visual images in combination provide a virtually unambiguous compilation of evidence that normal spiral galaxies can contain intrinsic redshifts at least as large as 5000 km s$^{-1}$.  While large magnitude errors resulting in large TFR errors cannot be completely ruled out, the small uncertainty in the HyperLeda magnitudes, independent distances to companion galaxies, and the linear diameter analysis make the possibility of large magnitude errors extremely unlikely.  In the absence of large rotational velocity errors, the evidence from HyperLeda images also makes the possibility of large magnitude errors doubtful.

If a Hubble constant lower than the preferred H$_0$=72 km s$^{-1}$ Mpc$^{-1}$ is adopted the intrinsic redshifts would be even larger than identified here and in paper 1.  The implications of this result present tremendous challenges for understanding the nature of extragalactic redshifts.

*Acknowledgements: This research has made use of the HyperLeda database (Paturel et al 2003) compiled by the LEDA team at the CRAL-Observatoire de Lyon (France).  I would like to thank the referee for comments that helped provide significant improvements in the description of these results.*

**References:**

Arp, H. 1988, Astron. Astrophys. 202, 70
Arp, H. 1990, Astrophys. Space Sci. 167, 183
Bell, M.B. 2002, Astrophys. J. 566, 705
Bell, M.B. 2004, Astrophys. J. accepted, astroph/0409025
Drozdovsky, I. and Karachentsev, I.  2000, Astron. Astrophys. Suppl. 142,425
Freedman, W. et al 2001, Astrophys. J. 553, 47
Karachentsev, I., et al 2003, Astron. Astrophys. 404,93
Mathewson, D. & Ford, V. 1996, Astrophys. J. Suppl. 107,97
Newman, J., Zepf, S., Davis, M., Freedman, W., Madore, B., et al: 1999 Astrophys. J. 523, 506
Paturel, G., Theureau, G., Botinelli, L., Gouguenheim, L. et al 2003, Astron. Astrophys. 412,57
Prugniel, P. and Simien, F., 1996, Astron. Astrophys. 309,749
Russell, D.G. 2004, Astrophys. J. 607, 241
Russell, D.G. 2005, Astrophys. Space Sci. in press, astro-ph/0408348 (paper 1)
Tonry,  J.L. et al 2001, Astrophys. J. 546,681